\documentclass{iau} 

\usepackage{graphicx}

\usepackage[modulo, switch]{lineno}

\newcommand\degr{\hbox{$^\circ$}}

\newcommand\arcsec{\hbox{$^{\prime\prime}$}}

\title[Flows along arch filaments]{Flows along arch filaments observed in the 
    GRIS `very fast spectroscopic mode'}
 
\author[Gonz{\'a}lez Manrique et al.]{
    S.J.\ Gonz{\'a}lez Manrique$^{1,2}$,
    C.\ Denker$^1$,
    C.\ Kuckein$^1$,
    A.\ Pastor Yabar$^{3,4}$, 
    M.\ Collados$^3$, 
    M.\ Verma$^1$,
    H.\ Balthasar$^1$,
    A.\ Diercke$^{1,2}$,\\
    C.E.\ Fischer$^5$,
    P.\ G{\"o}m{\"o}ry$^6$,
    N.\ Bello Gonz{\'a}lez$^5$,
    R.\ Schlichenmaier$^5$, 
    M.\ Cubas Armas$^{3,4}$,
    T.\ Berkefeld$^5$,
    A.\ Feller$^5$,
    S.\ Hoch$^7$,\\
    A.\ Hofmann$^1$,
    A.\ Lagg$^7$,
    H.\ Nicklas$^8$,
    D. Orozco Su{\'a}rez$^{3,4}$,\\
    D.\ Schmidt$^9$,
    W.\ Schmidt$^5$,
    M.\ Sigwarth$^5$,
    M.\ Sobotka$^{10}$,\\
    S.K.\ Solanki$^{7,11}$,
    D.\ Soltau$^5$,
    J.\ Staude$^1$,
    K.G.\ Strassmeier$^1$,\\
    R.\ Volkmer$^5$,
    O.\ von der L{\"u}he$^5$, \and
    T.\ Waldmann$^5$}
       
\affiliation{
    $^1$Leibniz-Institut f{\"u}r Astrophysik Potsdam, 14482 Potsdam, Germany\\
        email: \texttt{smanrique@aip.de}\\\smallskip
    $^2$Universit{\"a}t Potsdam, Institut f{\"u}r Physik and Astronomie, 14476 
        Potsdam, Germany\\\smallskip
    $^3$Instituto de Astrof{\'\i}sica de Canarias, 38205 La Laguna, Tenerife,
        Spain\\\smallskip
    $^4$Departamento de Astrof{\'\i}sica, Universidad de La Laguna, 38205 La 
        Laguna, Tenerife, Spain\\\smallskip
    $^5$Kiepenheuer-Institut f{\"u}r Sonnenphysik, 79104 Freiburg, 
        Germany\\\smallskip
    $^6$Astronomical Institute, Academy of Sciences, 05960 Tatransk{\'a} 
        Lomnica, Slovak Republic\\\smallskip
    $^7$Max-Planck-Institut f{\"u}r Sonnensystemforschung, 37077 
        G{\"o}ttingen, Germany\\\smallskip
    $^8$Institut f\"ur Astrophysik, Georg-August-Universit\"at, 37077 
        G\"ottingen, Germany\\\smallskip
    $^{9}$National Solar Observatory, Sacramento Peak, Sunspot, NM 88349, 
        USA\\\smallskip
    $^{10}$Astronomical Institute, Academy of Sciences, 25165 Ond\v{r}ejov, 
        Czech Republic\\\smallskip
    $^{11}$School of Space Research, Kyung Hee University, Yongin, 
        Gyeonggi-Do, 446-701, Korea}

\pubyear{2017}
\volume{327}
\jname{Fine Structure and Dynamics of the Solar Atmosphere}
\editors{S.\ Vargas Dom{\'\i}nguez, A.G.\ Kosovichev, L.\ Harra, P.\ Antolin, 
    eds.}

\begin{document}

\maketitle


\begin{abstract}
A new generation of solar instruments provides improved spectral, spatial, 
and temporal resolution, thus facilitating a better understanding of dynamic 
processes on the Sun. High-resolution observations often reveal 
multiple-component spectral line profiles, e.g., in the near-infrared 
He\,\textsc{i} 10830~\AA\ triplet, which provides information about the 
chromospheric velocity and magnetic fine structure. We observed an emerging 
flux region, including two small pores and an arch filament system,
on 2015 April~17 with the `very fast spectroscopic mode' of the GREGOR
Infrared Spectrograph (GRIS) situated at the 1.5-meter GREGOR solar telescope 
at Observatorio del Teide, Tenerife, Spain. We discuss this method of obtaining 
fast (one per minute) spectral scans of the solar surface and its potential 
to follow dynamic processes on the Sun. We demonstrate the performance of 
the `very fast spectroscopic mode' by tracking chromospheric high-velocity features in the arch filament
system.
\end{abstract}

%
%

\section{Introduction}

The emergence of magnetic flux \cite[(Zwaan 1985)]{Zwaan1985} in the form of
rising $\Omega$-loops is characterized by strong photospheric and chromospheric
downflows near footpoints, coalescence of small-scale magnetic features, 
and dark fibrils within the arch filament system (AFS) with upward motions connecting opposite-polarity
patches as well as abnormal granulation and transverse magnetic fields between
these patches. The horizontal proper motions were studied, for example, by
\cite{Strous1996} and \cite{Verma2016} using feature and local correlation
tracking \cite[(LCT, Verma \& Denker 2011)]{Verma2011}. Investigating strong
chromospheric absorption lines, e.g., the He\,\textsc{i} 10830~\AA\ triplet
reveals supersonic downflows near the footpoints of $\Omega$-loops
\cite[(e.g., Schmidt \& Schlichenmaier 2000, Lagg \etal\ 2007, Xu \etal\ 2010, Balthasar \etal\ 2016, 
Gonz\'alez Manrique \etal\ 2016)]{Lagg2007, Xu2010, Balthasar2016,
GonzalezManrique2016}. The size of emerging flux regions (EFRs) reaches from small sunspots, over 
pores, to micro-pores \cite[(Gonz\'alez Manrique \etal\
2017)]{GonzalezManrique2017}, which all exhibit similar properties, even though 
the velocities scale with size. Besides line fitting a variety of spectral
inversion methods were employed in these studies to infer photospheric and
chromospheric velocities.

At the GREGOR telescope, GRIS \cite[(Collados \etal\ 2012)]{Collados2012} in 
the `very fast spectroscopic mode' (VFSM) achieves a one-minute cadence comparable to that of imaging
spectrometers \cite[(Denker 2010)]{Denker2010a}, covering an area of about $66\arcsec \times 24\arcsec$.  Even higher cadences become possible by sacrificing spatial 
and spectral resolution. For example, the Fast Imaging Solar Spectrograph
\cite[(FISS, Chae \etal\ 2013)]{Chae2013} is a dual-beam \'echelle 
spectrograph installed at the 1.6-meter New Solar Telescope (NST) at Big 
Bear Solar Observatory, California. It scans a field-of-view (FOV) of 
$60\arcsec \times 60\arcsec$ in less than 20~s, either in the H$\alpha$ 
line at 6562.8~\AA\ or in the Ca\,\textsc{ii} NIR line at 8542.1~\AA. 
Conceptually similar to GRIS but designed as a multi-slit, dual-beam
spectropolarimeter, the Facility Infrared Spectropolarimeter \cite[(FIRS, 
Jaeggli \etal\ 2010)]{Jaeggli2010} shortens the scanning time by recording
spectra with four slits at the same time. FIRS is operated at the 0.76-meter 
Dunn Solar Telescope (DST) at the National Solar Observatory/Sacramento Peak, 
New Mexico. Both FISS and FIRS share many design concepts with GRIS, and our
motivation is to illustrate the potential of GRIS/VFSM in particular and
high-cadence, two-dimensional spectroscopy in general.


\section{Observations}\label{SEC2}

The EFR/AFS was observed with GRIS at the 1.5-meter GREGOR solar telescope 
\cite[(Denker \etal\ 2012, Kneer 2012, Schmidt \etal\ 2012)]{Denker2012,
Kneer2012, Schmidt2012} on 2015 April~17 as part of a coordinated observing 
campaign. \cite{GonzalezManrique2016} described preliminary results of 
fitting dual-component He\,\textsc{i} 10830~\AA\ spectral profiles. 
Complementary data were taken with the GREGOR Fabry-P\'erot Interferometer 
\cite[(GFPI, Puschmann \etal\ 2012)]{Puschmann2012} using the photospheric
Fe\,\textsc{i} 6302~\AA\ line and with the CRisp Imaging Spectro-Polarimeter
\cite[(CRISP, Scharmer \etal\ 2008)]{Scharmer2008} situated at the Swedish 
Solar Telescope \cite[(SST, Scharmer \etal\ 2003)]{Scharmer2003} at 
Observatorio del Roque de los Muchachos, La Palma, Spain using the
photospheric Fe\,\textsc{i} 6173~\AA\ and chromospheric Ca\,\textsc{ii} 
8542~\AA\ lines.

\begin{figure}[t]
\includegraphics[width=\textwidth]{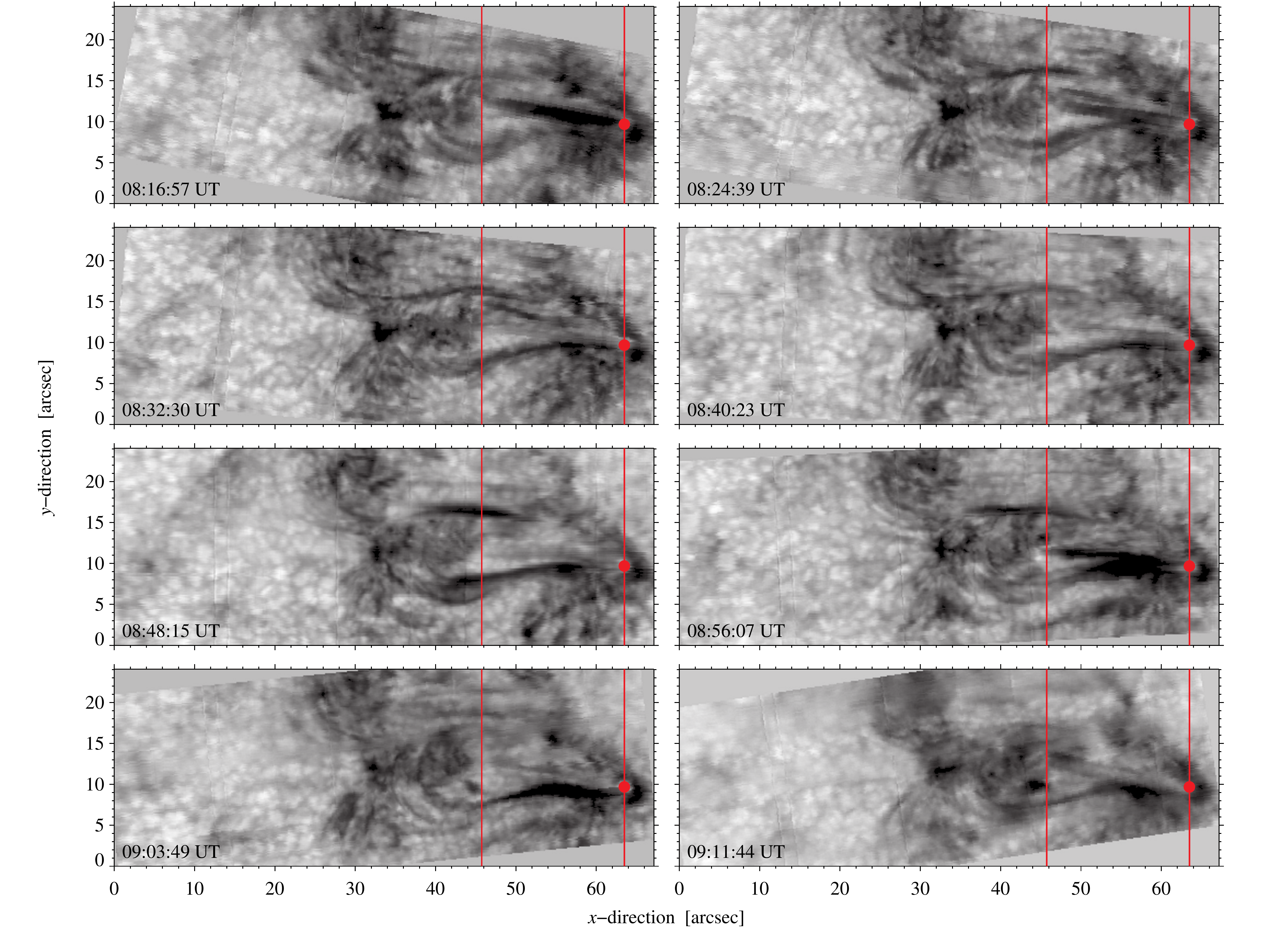}
\caption{Time-series of eight slit-reconstructed line-core intensity images 
    of the He\,\textsc{i} triplet's red component. The time interval between 
    successive images is about 8~min. Positions of the space-time diagrams
    (Figs.~\ref{FIG02} and \ref{FIG03}) are indicated by vertical lines, and 
    the filled circles refer to the location of a strong downflow kernel with 
    dual-component He\,\textsc{i} profiles (see Fig.~\ref{FIG04}).}
\label{FIG01}
\end{figure}

GRIS was operated in the spectroscopic mode covering an 
18-\AA-wide spectral region containing the photospheric Si\,\textsc{i} 
10827~\AA, Ca\,\textsc{i} 10834~\AA, and Ca\,\textsc{i} 10839~\AA\ lines as 
well as the chromospheric He\,\textsc{i} triplet at 10830~\AA. The dispersion
was 18.03~m\AA\ pixel$^{-1}$. The pixel size along the slit and spatial step
size were almost identical, i.e., 0.136\arcsec\ and 0.134\arcsec, respectively.
Thus, the spectroscopic scans with 180 steps covered a FOV of
about $66.3\arcsec \times 24.1\arcsec$. A single 
exposure of 100~ms duration was recorded at each spatial step, and it took about
58~s to complete the scan. The observing sequence
continued for about one hour and resulted in 64 data cubes. The 
GREGOR Adaptive Optics System \cite[(GAOS, Berkefeld \etal\ 
2012)]{Berkefeld2012} provided real-time image correction. The rotation compensation 
of the alt-azimuthal
mount \cite[(Volkmer \etal\ 2012)]{Volkmer2012} was not yet available so that
the FOV turned by 22.3\degr, which was corrected in post-processing, whereas 
the image rotation of 0.4\degr\ in one map was negligible.


\section{Results}\label{SEC3}

The fast cadence of GRIS/VFSM offers access to the highly dynamic chromosphere,
which is monitored in this study using the red component of the He\,\textsc{i}
10830~\AA\ triplet. The time-series of slit-reconstructed He\,\textsc{i}
line-core intensity images in Fig.~\ref{FIG01} affirms the good to very 
good seeing conditions and demonstrates the stable performance of the AO 
system. Moments of mediocre seeing conditions are rare but become more 
frequent towards the end of the time-series. They appear in the line-core
intensity maps as elongated stripes with low contrast and correspond typically
to time intervals of about 10~s. However, for most of the time, the granulation
pattern and chromospheric He\,\textsc{i} absorption features with sub-arcsecond
fine structure are very prominently visible.

Spectral scan No.~33 at 08:48:15~UT (fifth panel in Fig.~\ref{FIG01}) serves 
as reference for alignment and rotation correction of all spectral data cubes. 
Light gray areas indicate missing information in the region-of-interest 
(ROI), which contains two pores of opposite magnetic polarity, one in the 
center of the ROI at the AO lock-point and the other one at the right edge of 
the ROI. New magnetic flux is still emerging, which leads to an AFS connecting 
both pores and other small-scale magnetic elements with multiple dark filament
strands. Individual arch filaments change significantly on time-scales of minutes or
less, while the overall morphology of the EFR/AFS remains fairly constant over
the observing period. Time-lapse movies reveal motions along the fibrils, which
are consistent with continuously emerging $\Omega$-loops, where cool material
rises to chromospheric heights and beyond at the loop tops and then drains
towards the footpoints of the loops.

\begin{figure}[t]
\includegraphics[width=0.495\textwidth]{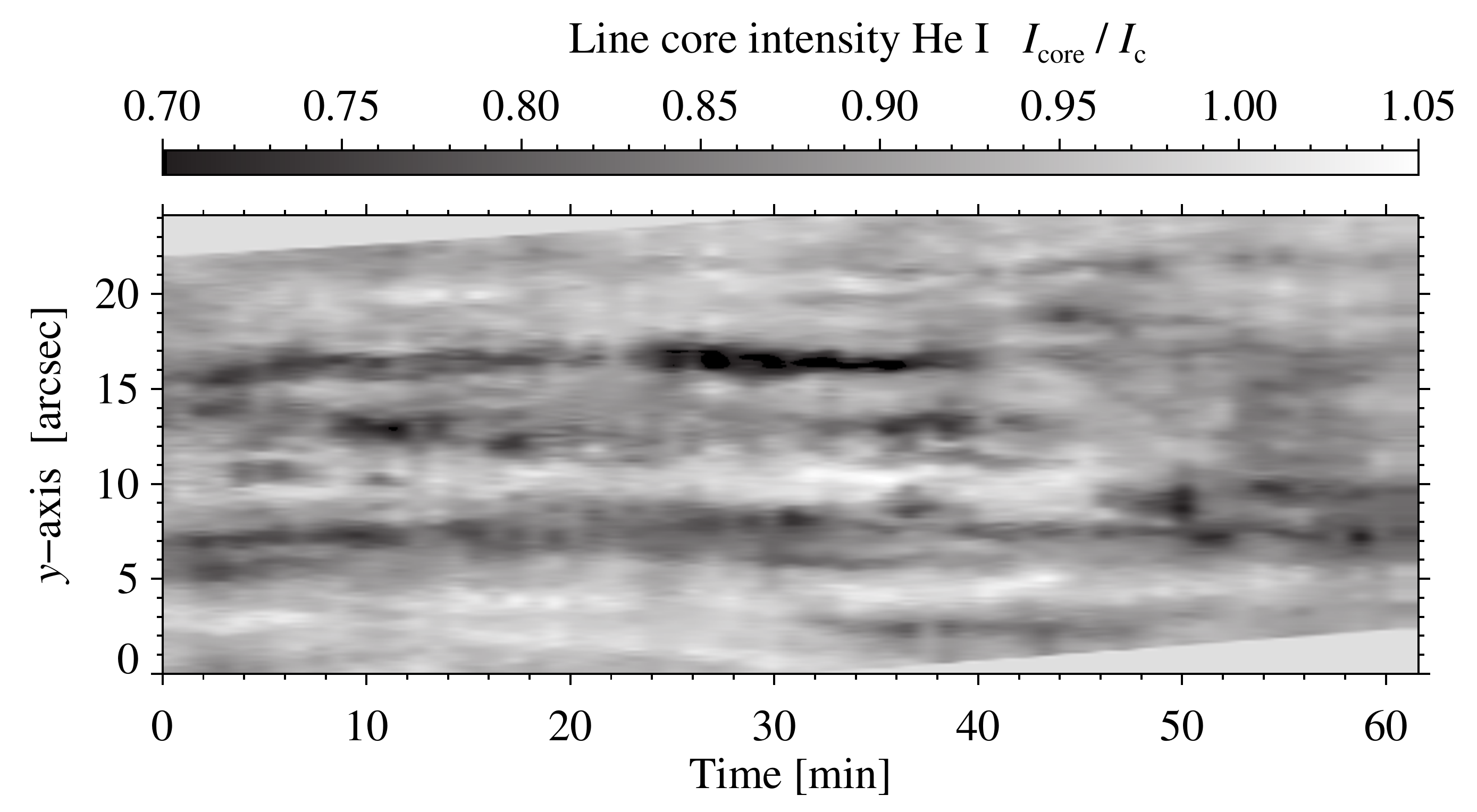}
\hfill
\includegraphics[width=0.495\textwidth]{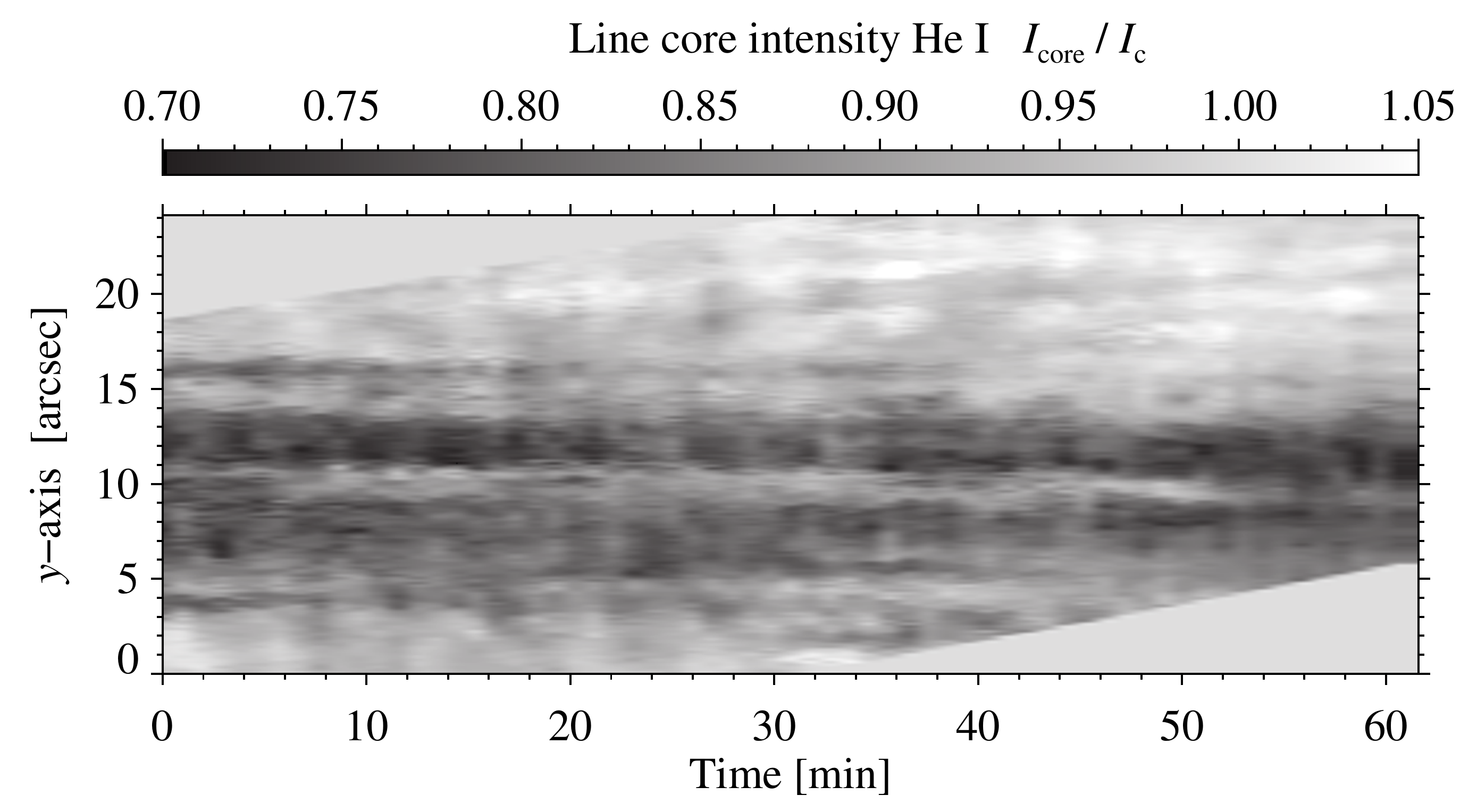}
\caption{Space-time diagram of the He\,\textsc{i} line-core intensity
    $I_\mathrm{core}$ (normalized to the continuum intensity $I_\mathrm{c}$)
    taken at the footpoint (\textit{right}) and loop top (\textit{left}) of 
    the AFS. The positions are marked in Fig.~\ref{FIG01} by vertical lines.}
\label{FIG02}
\end{figure}

\begin{figure}[t]
\includegraphics[width=0.495\textwidth]{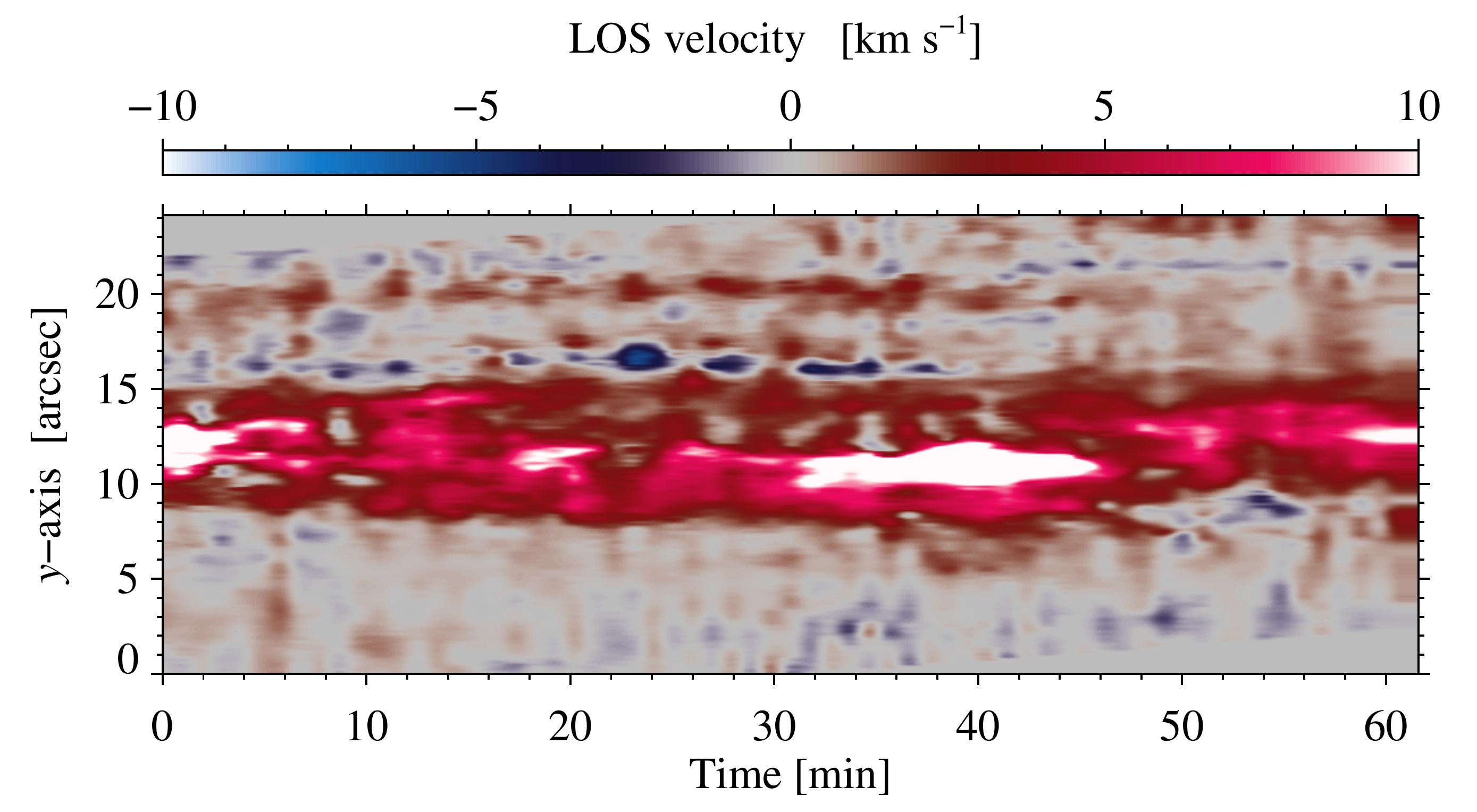}
\hfill
\includegraphics[width=0.495\textwidth]{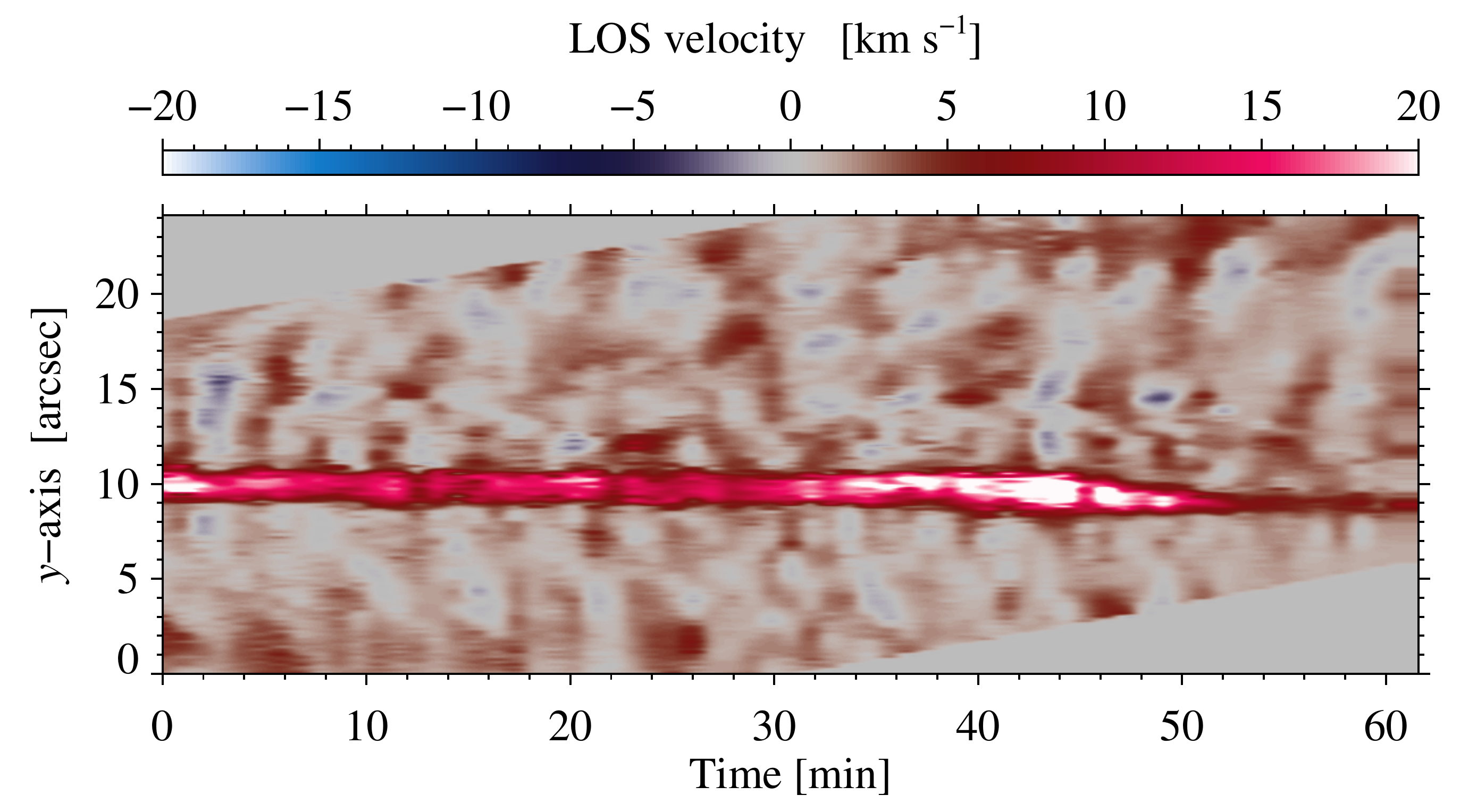}
\caption{Space-time diagram of the He\,\textsc{i} LOS velocity 
    (single-component fit) taken at the footpoint (\textit{right}) and loop top 
    (\textit{left}) of the AFS. The positions are marked in Fig.~\ref{FIG01} 
    by vertical lines.}
\label{FIG03}
\end{figure}

The most interesting features in the ROI are the loop tops and footpoints
of the arch filaments. Two vertical lines in all panels of
Fig.~\ref{FIG01} mark the positions, where space-time diagrams display 
the temporal evolution of the He\,\textsc{i} line-core intensity
and line-of-sight (LOS) velocity in Figs.~\ref{FIG02} and \ref{FIG03},
respectively. Again, light gray areas represent missing values. The majority 
of LOS velocities results from single-component fits to the observed line 
profiles unless dual-component fits are more appropriate. In this case, the
LOS velocity refers to the fast component. The left panel 
of Fig.~\ref{FIG02} tracks the rising loop top of a thin dark fibril $(y \approx
17\arcsec)$, which becomes darkest about 30~min after the start of the
time-series and abruptly vanishes after just over 40~min. A second, more
persistent footpoint of a different loop exists at $y \approx 8\arcsec$, which
shows more substructure streaming rapidly past the crosswise line. The 
associated LOS velocities (left panel of Fig.~\ref{FIG03}) of the former fibril
show upflows approaching 5~km~s$^{-1}$, which are typical for a rising 
$\Omega$-loop, whereas the latter absorption feature near the footpoint of a 
loop exhibits strong variations of downflows in excess of 10~km~s$^{-1}$. The
right panels of Figs.~\ref{FIG02} and~\ref{FIG03} are dedicated to a small 
kernel with a diameter of 2\,--\,3\arcsec, where strong persistent downflows 
are present exceeding 20~km~s$^{-1}$, sometimes even reaching more than
40~km~s$^{-1}$. These downflows intensify just before the kernel vanishes at 
$t \approx 50$~min. At this location He\,\textsc{i} absorption is moderately
strong and does not change much in time. This downflow kernel is pointed out 
by filled circles with diameters of about 1.4\arcsec\ at coordinates
(63.2\arcsec, 9.6\arcsec).

\begin{figure}[t]
\centering
\includegraphics[width=0.8\textwidth]{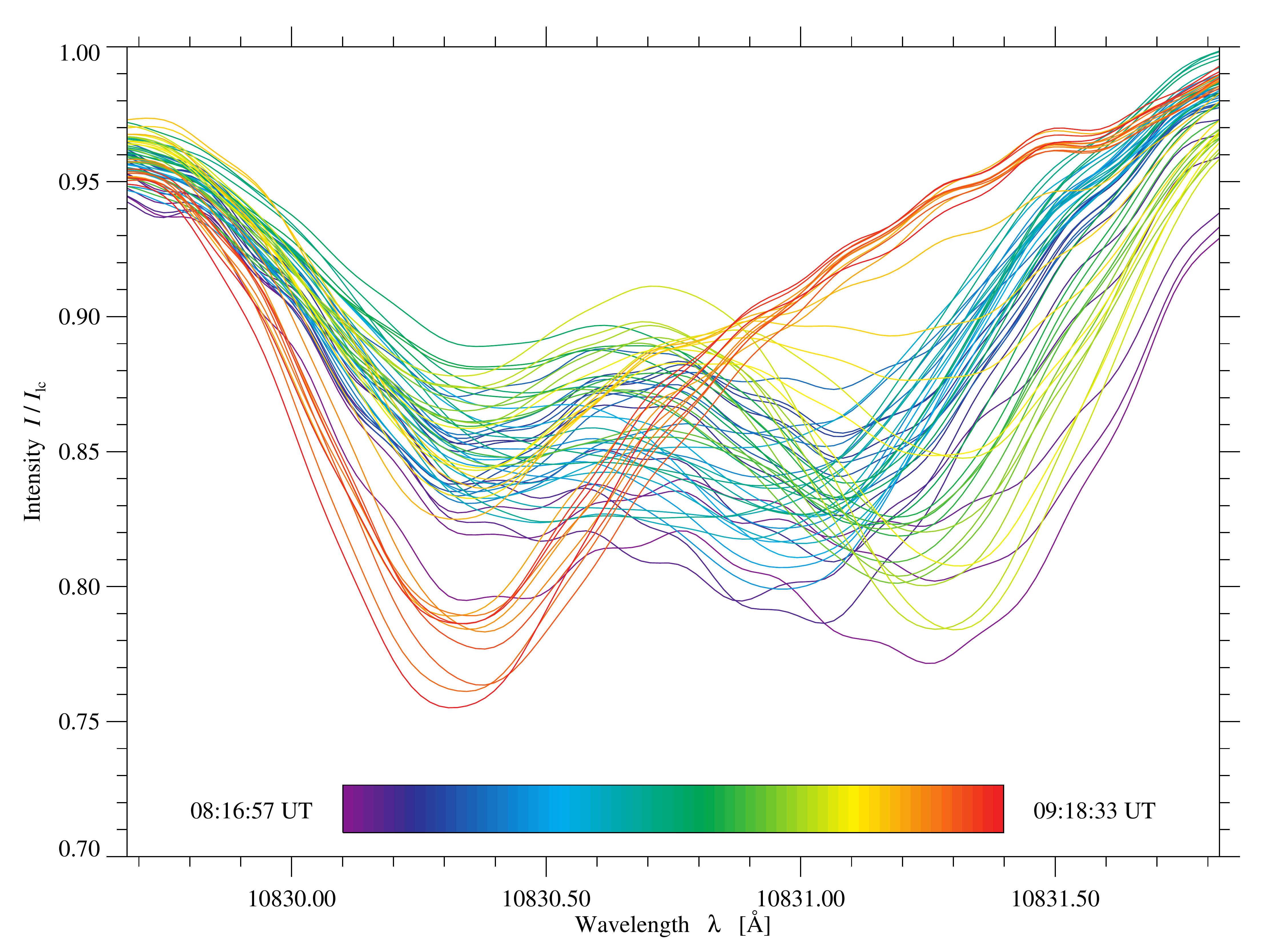}
\caption{Temporal evolution of spectral profiles (normalized to the local
    continuum intensity $I_\mathrm{lc}$) for the red component of the
    He\,\textsc{i} triplet. The rainbow-colored bar marks the elapsed time after
    the start of the one-hour time-series. The spectra were taken at a strong
    downflow kernel marked by filled circles in Fig.~\ref{FIG01}. The spectra
    were slightly smoothed to avoid a cluttered display.}
\label{FIG04}
\end{figure}

The temporal evolution of He\,\textsc{i} spectra within this downflow kernel
deserves further scrutiny. In Fig.~\ref{FIG04}, the time since the start of 
the time-series is color coded. The displayed spectra arise from 69 locally
averaged spectra contained within the filled circles. They were slightly 
smoothed for better display but still contain some high-frequency oscillations
due to fringes in the NIR spectra. The region was specifically selected because 
of the strong downflows and the presence of dual-component He\,\textsc{i} 
spectral profiles. The rainbow color code immediately reveals time intervals,
where the shape of the red component significantly changes. Initially, the 
slow and fast component form \textsf{w}-shaped profiles with well separated
minima of roughly the same intensity. Their line-core intensities are in the
range $I_\mathrm{core} / I_\mathrm{c} \approx 0.80$\,--\,0.85. At around
08:41~UT, the combined profile of the slow and fast component becomes more
`bathtub'-like, i.e., the minima are no longer well separated. Starting at
08:50~UT, the fast component becomes dominant reaching line-core intensities of
$I_\mathrm{core} / I_\mathrm{c} < 0.8$, and the difference in line-core intensity
between the slow and fast component can be as large as $\Delta I \approx 0.1$.
The transition to single-component profiles begins abruptly at 09:08~UT 
resulting in profiles with extended wings towards higher wavelengths and in deep line cores of the slow component.


\section{Discussion and conclusions}\label{SEC4}

The observed supersonic downflows (up to 40~km~s$^{-1}$) at the footpoints of
arch filaments near pores of an EFR are in very good agreement with previous
observations by, for example, \cite{Lagg2007} and \cite{Xu2010}. However,
magnetic field information is obviously missing in VFSM but can be provided 
in the future with polarimetric GFPI observations \cite[(Balthasar \etal\
2011)]{{Balthasar2011}}. The downflow kernel discussed above has a size of 
just a few seconds of arc and is located in close proximity to the pore.
Dual-component He\,\textsc{i} profiles with a strong fast component are
continuously present in the vicinity of both pores. They are also 
sporadically encountered at other locations along the dark fibrils of 
the AFS. The high-cadence GRIS observations reveal significant changes 
of the spectral line shapes on time-scales of a few minutes inside the 
downflow kernels. 

The VFSM of GRIS reaches a one-minute cadence comparable to imaging 
spectrometers by removing the dual-beam polarimeter, which increases the 
photon flux at the detector by a factor of two. Recording just intensity 
spectra and not their polarization state, lowers the scanning time by a
factor of four. Finally, not accumulating spectra at each scan position,
another factor of 8\,--\,10
can be gained. Thus, GRIS/VFSM
operations mirror the FISS performance and are also applicable to FIRS.

We introduced various ways of visualizing the dynamics of EFRs/AFSs: 
serial maps of physical parameters, space-time diagrams, and sequential spectra
of downflow kernels with dual-component He\,\textsc{i} profiles. We also
attempted applying LCT to a sequence of He\,\textsc{i} line-core intensity maps.
However, the fibril structure of the AFS evolves too quickly to be captured 
inside the sampling window but tracking slower photospheric motions is possible.
This article mainly deals with methodical  and technical aspects of the VFSM, 
the detailed scientific analysis is deferred to a forthcoming article, where 
we will  investigate, for example, if the plasma contained in arch filaments,
which exhibit supersonic LOS downflows near the footpoints of the AFS, 
reaches the photosphere.

%
%

\acknowledgements The 1.5-meter GREGOR solar telescope was built by a German 
consortium under the leadership of the Kiepenheuer-Institut f\"ur Sonnenphysik 
in Freiburg with the Leibniz-Institut f\"ur Astrophysik Potsdam (AIP), the 
Institut f\"ur Astrophysik G\"ottingen, and the Max-Planck-Institut f\"ur 
Sonnensystemforschung in G\"ottingen as partners, and with contributions by the 
Instituto de Astrof\'{\i}sica de Canarias and the Astronomical Institute of the 
Academy of Sciences of the Czech Republic. SJGM is supported by the Leibniz
Graduate School for Quantitative Spectroscopy in Astrophysics. He is also grateful for an
 IAU travel grant enabling him to attend the symposium.  
This study is supported by the European Commission's FP7 Capacities Program under 
Grant Agreement No.~312495.



\begin{thebibliography}{}
\bibitem[Balthasar \etal\ (2011)]{Balthasar2011}
    Balthasar, H., Bello Gonz\'alez, N., Collados, M., \etal\ 2011,
    \textit{ASP-CS}, 437, 351
\bibitem[Balthasar \etal\ (2016)]{Balthasar2016}
    Balthasar, H., G{\"o}m{\"o}ry, P., Gonz{\'a}lez Manrique, S.J., \etal\ 2016,
    \textit{AN}, 337, 1050
\bibitem[Berkefeld \etal\ (2012)]{Berkefeld2012}   
    Berkefeld , T., Schmidt, D., Soltau, D., \etal\ 2012, 
    \textit{AN}, 333, 863
\bibitem[Chae \etal\ (2013)]{Chae2013}
    Chae, J., Park, H.-M., Ahn, K., \etal\ 2013, 
    \textit{Solar Phys.}, 288, 1
\bibitem[Collados \etal\ (2012)]{Collados2012}
    Collados, M., L{\'o}pez, R., P{\'a}ez, E., \etal\ 2012, 
    \textit{AN}, 333, 872
\bibitem[Denker (2010)]{Denker2010a}   
    Denker, C.\ 2010, 
    \textit{AN}, 331, 648
\bibitem[Denker \etal\ (2012)]{Denker2012}  
    Denker, C., von der L{\"u}he, O., Feller, A., \etal\ 2012, 
    \textit{AN}, 333, 810
\bibitem[Gonz{\'a}lez Manrique \etal\ (2016)]{GonzalezManrique2016}
    Gonz{\'a}lez Manrique, S.J., Kuckein, C., Pastor Yabar, A., \etal\ 2016, 
    \textit{AN}, 337, 1057
\bibitem[Gonz{\'a}lez Manrique \etal\ (2017)]{GonzalezManrique2017}    
    Gonz{\'a}lez Manrique, S.J., Bello Gonz\'alez, N., Denker, C.\ 2017,
    \textit{A\&A}, accepted
\bibitem[Jaeggli \etal\ (2010)]{Jaeggli2010}
    Jaeggli, S.A., Lin, H., Mickey, D.L., \etal\ 2010, 
    \textit{MemSAIt}, 81, 763
\bibitem[Kneer (2012)]{Kneer2012}   
    Kneer, F. 2012, 
    \textit{AN}, 333, 790
\bibitem[Lagg \etal\ (2007)]{Lagg2007}      
    Lagg, A., Woch, J., Solanki, S.K., Krupp, N.\ 2007,
	\textit{A\&A}, 462, 1147
\bibitem[Puschmann \etal\ (2012)]{Puschmann2012}   
    Puschmann, K.G., Denker, C., Kneer, F., \etal\ 2012, 
    \textit{AN}, 333, 880
\bibitem[Scharmer \etal\ (2003)]{Scharmer2003}    
    Scharmer, G.B., Bjelksjo, K., Korhonen, T.K., \etal\ 2003,
    \textit{Proc. SPIE}, 4853, 341
\bibitem[Scharmer \etal\ (2008)]{Scharmer2008}   
    Scharmer, G.B., Narayan, G., Hillberg, T., \etal\ 2008, 
    \textit{ApJ}, 689, L69
\bibitem[Schmidt \etal\ (2012)]{Schmidt2012}    
    Schmidt, W., von der L{\"u}he, O., Volkmer, R., \etal\ 2012, 
    \textit{AN}, 333, 796
\bibitem[Schmidt \& Schlichenmaier (2000)]{Schmidt2000}    
    Schmidt, W., Schlichenmaier, R.\ 2000, 
    \textit{A\&A}, 364, 829    
\bibitem[Strous \etal\ (1996)]{Strous1996}     
    Strous, L., Scharmer, G., Tarbell, T.D., \etal\ 1996,
    \textit{A\&A}, 306, 947
\bibitem[Verma \etal\ (2016)]{Verma2016}   
    Verma, M., Denker, C., Balthasar, H., Kuckein, C., \etal\ 2016,
    \textit{A\&A}, 596, A3
\bibitem[Verma \& Denker (2011)]{Verma2011}   
    Verma, M., Denker, C.\ 2011, 
    \textit{A\&A}, 529, A153
\bibitem[Volkmer \etal\ (2012)]{Volkmer2012}   
    Volkmer, R., Eisentr{\"a}ger, P., Emde, P., \etal\ 2012, 
    \textit{AN}, 333, 816
\bibitem[Xu \etal\ (2010)]{Xu2010}   
    Xu, Z., Lagg, A., Solanki, S.K.\ 2010,
    \textit{A\&A}, 520, A77
\bibitem[Zwaan (1985)]{Zwaan1985}     
    Zwaan, C.\ 1985,
    \textit{Solar Phys.}, 100, 397
\end{thebibliography}
\end{document}